\documentclass{article}

 \usepackage[creativeai, final]{neurips_2025}

\usepackage[utf8]{inputenc} 
\usepackage[T1]{fontenc}    
\usepackage{hyperref}       
\usepackage{url}            
\usepackage{booktabs}       
\usepackage{amsfonts}       
\usepackage{nicefrac}       
\usepackage{microtype}      
\usepackage{xcolor}         
\usepackage{wrapfig}
\usepackage{graphicx}

\usepackage{subfig}

\title{``Jutters''}

\author{
  Meike Driessen$^{*}$\\
  \texttt{meike-driessen@hotmail.com} \\
  \And
  Selina Khan$^{*}$\\
  University of Amsterdam \\
  \texttt{s.j.khan@uva.nl} \\
  \And
  Gonçalo Marcelino$^{*}$\\
  University of Amsterdam \\
  \texttt{g.barretoferreiramarcelino@uva.nl} \\
}

\newcommand\freefootnote[1]{%
  \let\thefootnote\relax%
  \footnotetext{#1}%
  \let\thefootnote\svthefootnote%
}

\begin{document}

\maketitle
\freefootnote{$^*$ Authors with equal contribution.}

\begin{abstract}
This project explores how we engage with AI-generated content through the lens of the \textit{jutter}: Dutch coastal foragers who comb the shoreline after storms, gathering and repurposing what the sea leaves behind. Reflecting how our lives are increasingly shaped by AI-generated media, we create a beach-like installation that blends real shoreline debris with AI-transformed images and videos. Visitors are invited to explore this space as contemporary \textit{jutters}, deciding what to keep and what to discard. In doing so, the project reimagines AI-imagery as material for reflection, encouraging a more discerning engagement with the content that drifts through our feeds. A video preview of the installation can be found at \small\url{https://www.youtube.com/watch?v=L6319Ii7MT8}.
\end{abstract}

\begin{figure}[!h]
    \centering
\includegraphics[width=\linewidth]{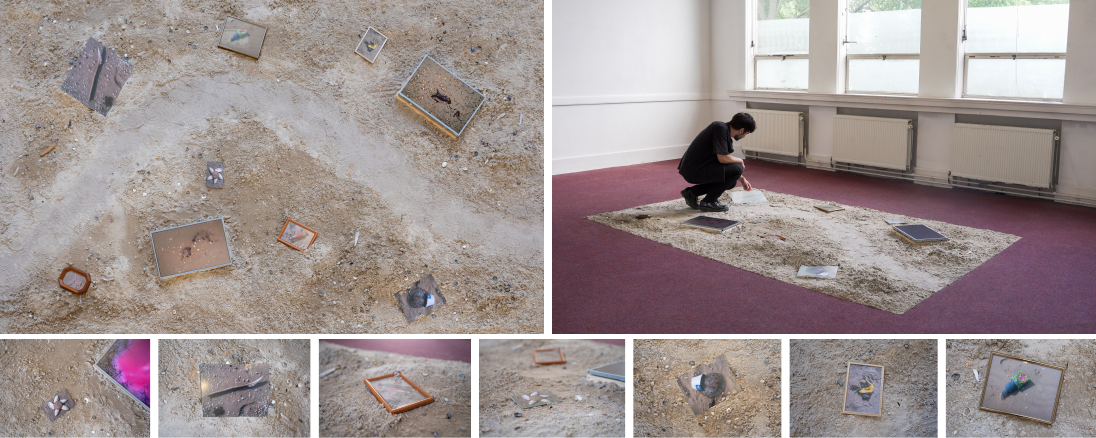}
    \caption{Overview of installation. Visitors are invited to become \textit{jutters} and walk the sandy path to explore the artificial beach, deciding which items are worth keeping.}
    \label{fig:cover}
\end{figure}

\section*{Background}
For the longest time, in the early morning hours after a storm, people could be seen wandering the windy, grey beaches of the Netherlands. They searched for objects washed ashore, items transformed by the waves, often broken, rinsed of their history, ownership, and intended use. The Dutch language has a name for these people: \textit{jutters}. \textit{Jutters} were usually poor people who resided near coastal areas, primarily living off possessions offered to them by the sea. Washed-up cargo, the wood from sunken ships, trinkets fallen overboard. These would be transformed into houses, furniture, new tools, or sold for money. Things once adrift were recontextualised, given new meaning by a tradition of curation born out of the struggle to survive \cite{jutters}.

Today, our messages in a bottle, our missing cargo, the things we send out, lose, or leave behind, are not only physical objects, they are also digital. They are the videos from our childhood posted to YouTube, the arguments we had in now-dead internet forums, the social media profiles we no longer have access to. Intelligent algorithms ingest and transform these digital footprints beyond recognition, and the resulting artefacts wash up on our feeds.

The majority of the content we consume digitally is no longer content we seek. It is content that is served to us, and a large part of this content is now AI-generated \cite{too_much_ai}. We are having an increasingly harder time distinguishing what is original, what is real, what is interesting, and what is purposeless~\cite{misinfo, lu2023seeing}. 

When we encounter an object, we engage with it not as an isolated entity but as a network of interconnected possibilities and relations \cite{harman2011tool}. It is perhaps not surprising that we are struggling to understand our relationship with AI-generated content. Like remnants of past lives washed ashore, AI-generated content lacks clear concepts of authorship, history, ownership, and intended use. The relations and possibilities that we are used to finding in things are fundamentally different or entirely absent when dealing with AI-generated content.

We aim to reframe these novel problems through an old perspective, that of the \textit{jutter} tradition. Through this process, we hope to discover more thoughtful and effective ways of curating our digital world and unveil new networks of possibilities and relations regarding AI-generated content. With this goal in mind, we created an installation where digital images of AI-generated objects are relocated to a physical space, a small artificial beach. Participants are then invited to engage with this space, and the objects within it, with the mindset of a \textit{jutter}. By walking through the sand and deciding, at their own pace, what is worth keeping and what can be left behind to be reclaimed by the sea, participants are encouraged to embody a crucial new type of curatorial role emerging in the age of AI-generated content.

\section*{Description of the installation}
The installation is composed of an approximately $2\times3$ meter space covered in sand. Some parts are dry, others humid, evoking a patch of beach near the sea. Lighter-coloured sand is used to define a path through which the participant is invited to walk, with darker-coloured sand used in the remaining areas of the installation. Shells, small pieces of driftwood, and other tidal debris collected from Noordwijk beach in the Netherlands, a site once frequented by \textit{jutters}, are scattered across the sand. Alongside these bits and pieces of a real beach are placed $7$ framed prints of AI-generated images, mirroring the objects encountered by \textit{jutters}, and $2$ screens displaying AI-transformed videos, mirroring the transformation process that an object undergoes when being hit by the waves and eroded by the sea.

As participants move through the installation, they are guided by an audio narration accompanied by the sound of waves. The narrator leads the participants through the space while introducing the \textit{jutter} tradition. The participants are invited to become \textit{jutters} of the digital sea themselves, and to reflect on the AI-transformed objects around them. The questions posed by the narrator prompt consideration of lost histories, altered meanings, and the new roles these digitally reshaped fragments might play in future contexts. Although the installation video is delivered in Dutch with English subtitles, the narration for the installation itself is translated into English.

\begin{figure}[!tbp]
  \centering
  \begin{minipage}[b]{0.48\textwidth}
    \includegraphics[width=\textwidth]{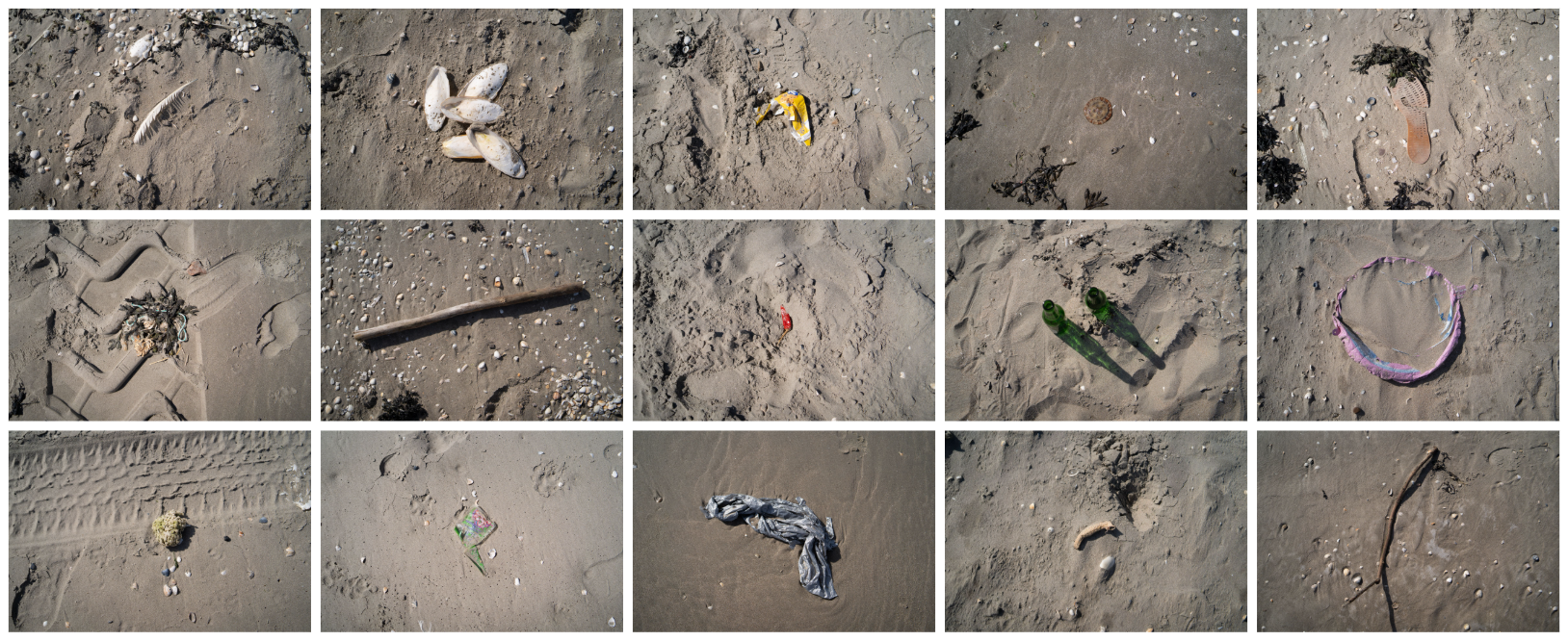}
    \caption{A sample of the over $100$ photographs of objects washed ashore taken at the beach.}
    \label{fig:original_images}
  \end{minipage}
  \hfill
  \begin{minipage}[b]{0.48\textwidth}
    \includegraphics[width=\textwidth]{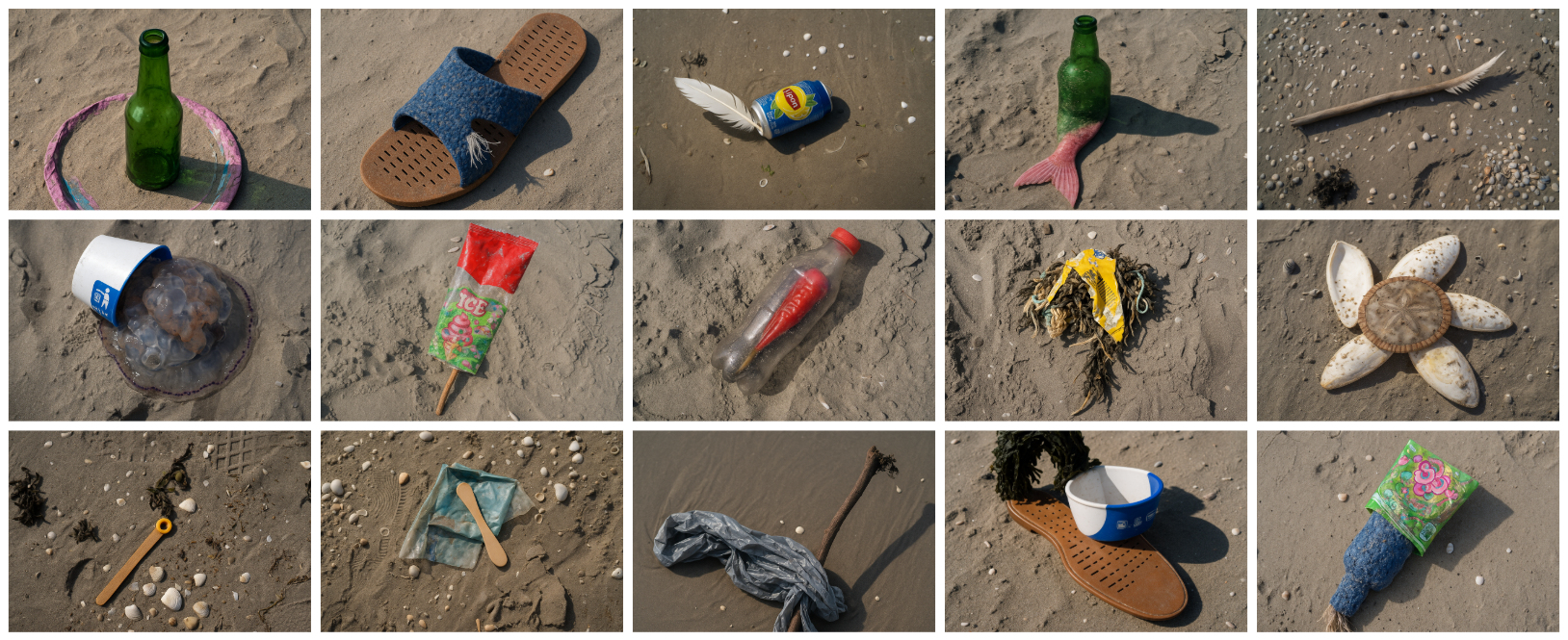}
    \caption{A sample of the AI images generated based on the photographs.}
    \label{fig:ai_images}
  \end{minipage}
\end{figure}

\subsubsection*{Technical details}
To create the AI-generated images, we initially photographed over $100$ objects found on the shore of Hoek van Holland beach, in the Netherlands (Figure~\ref{fig:original_images}). We then experimented with GPT-4 Omni \cite{openai2024gpt4ocard}, prompting the model to generate new images based on pairs or triplets of these photographs (Figure~\ref{fig:ai_images}). We selected images that appeared familiar but uncanny, evoking ambiguity and inviting interpretation. By placing these convincingly real-looking, yet unfamiliar objects on the artificial beach, we encouraged participants to engage in fresh acts of judgment, unanchored by prior associations.

To generate the videos, we recorded footage of objects found on the shore of Zandvoort beach in the Netherlands. We experimented with different video-to-video AI models, aiming to create videos of objects morphing while being rocked by the waves. The final videos were generated using WarpFusion \cite{Spirin2022}, an AI video-to-video transformation model which distorts form and motion in a surreal, but fluid style. Stable Diffusion \cite{Rombach_2022_CVPR} v1.4 was used as a baseline model, since we found that earlier Diffusion models generate more surreal imagery. For post-processing, the \textit{TopazLabs}\cite{topaz} "Frame Interpolation" tool was used to increase the frame rate of the videos. Finally, minor colouring adjustments were made manually using video editing software.

\section*{Acknowledgement of video source material}
The video preview incorporates imagery and audio from various external sources.
Archival footage was sourced from the publicly accessible collection of the Netherlands Institute for Sound and Vision \cite{bg1, bg2, bg3, bg4}, and images depicting \textit{jutters} were obtained from the Regional Archives of Alkmaar \cite{alkmaar1, alkmaar2}. The authors took additional photographs and videos at the \textit{Juttersmu-ZEE-um} museum, as well as in Zandvoort, Noordwijk, and Hoek van Holland. We thank the public archives for making these parts of their collections freely available, and thank the \textit{Juttersmu-ZEE-um} for facilitating on-site photography. We also thank the \textit{freesound.org} contributor “klankbeeld” for providing the background audio.

\section*{Biography of the authors}\
{\setlength\intextsep{0pt}
\begin{wrapfigure}{l}{25mm} 
    \includegraphics[width=1in,height=1in]{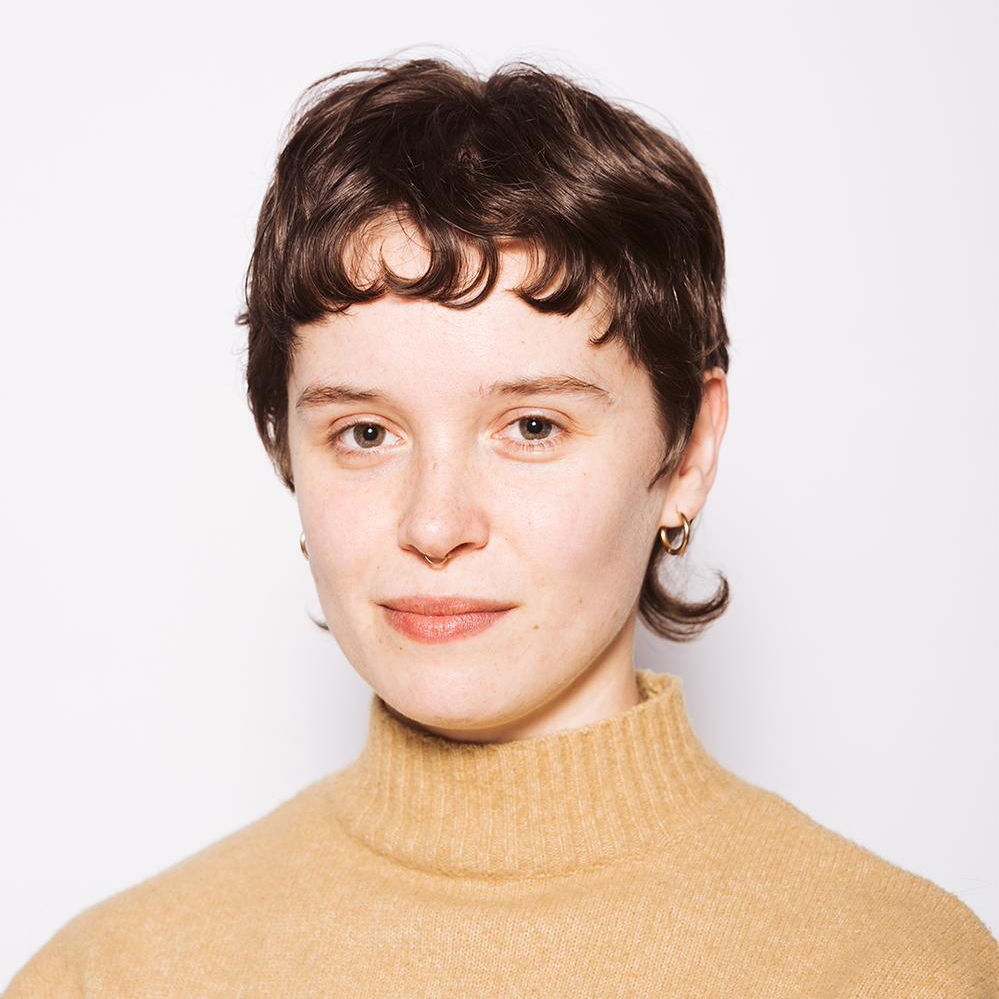}
\end{wrapfigure}
\noindent\textbf{Meike Driessen}\\ Meike is a visual artist and photographer. She graduated from St Joost School of Art \& Design in 2023. In her work \cite{meike2023vmdajk, meike2022bni, meike2023longing}, she explores themes of vulnerability, loneliness, and hope. With a strong fondness for poetic visual storytelling, Meike uses photography and film as subtle, interpretive languages. Rather than offering clear answers, Meike’s installations create quiet, immersive spaces where viewers can reflect, question, and connect on their terms. 
\par}
\par
{\setlength\intextsep{0pt}
\begin{wrapfigure}{l}{25mm} 
    \includegraphics[width=1in,height=1in]{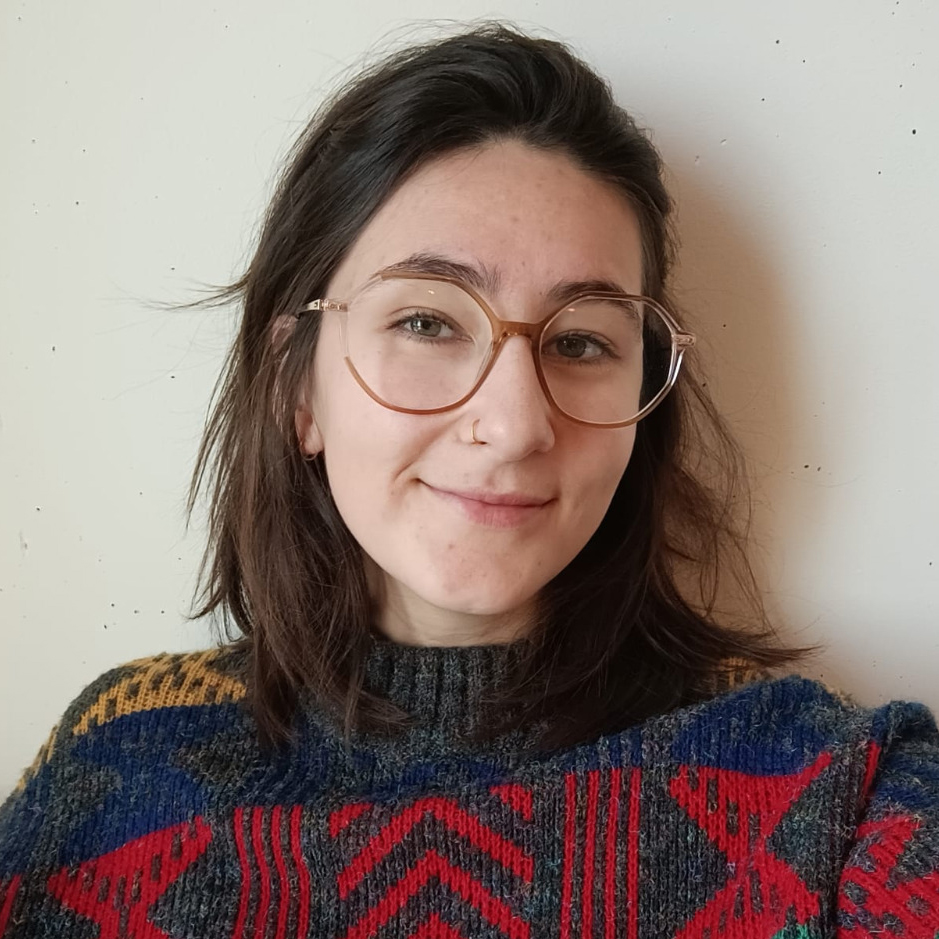}
\end{wrapfigure}\par
\noindent\textbf{Selina Khan}\\ Selina is a research assistant at the University of Amsterdam, focused on the intersection of AI and visual arts. Her research \cite{khan2021ukiyoe, khan2024civg} focuses on capturing nuanced, context-rich information from the artistic domain in AI systems and exploring how these technologies can deepen our understanding of art while respecting its subjectivity and cultural significance. Later this year, she will start her PhD on cultural bias in multimodal AI at the University of Amsterdam. \par}
\par
{\setlength\intextsep{0pt}
\begin{wrapfigure}{l}{25mm}     \includegraphics[width=1in,height=1in]{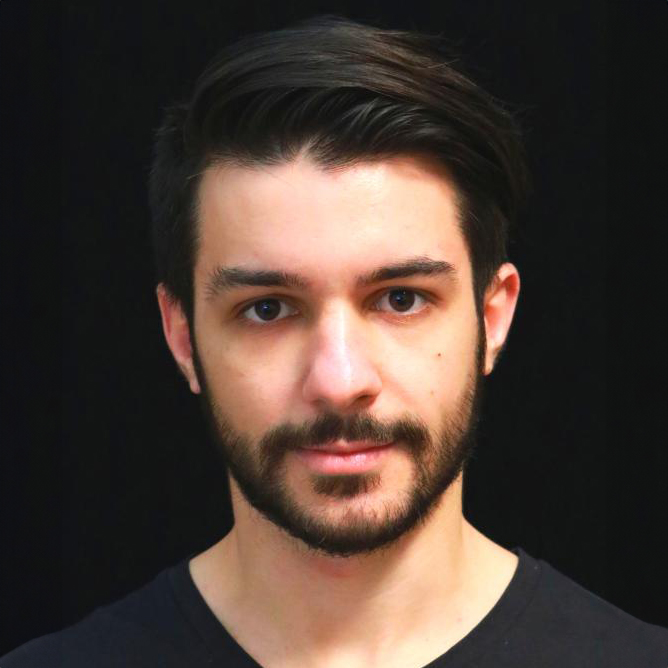}
\end{wrapfigure}\par
\noindent\textbf{Gonçalo Marcelino}\\Gonçalo is an AI researcher with a passion for art and philosophy. As part of his Master's thesis, he developed new AI methods for visual storytelling. His work led to several publications in top conferences \cite{marcelino2018ranking, marcelino2019benchmark, marcelino2021assisting} and received an award from the Fraunhofer Institute. He is currently a PhD candidate at the MultiX lab at the University of Amsterdam, collaborating with the Netherlands Forensic Institute to develop novel AI tools for combating money laundering.\par}

\pagebreak  
\newpage

\bibliography{references}

\end{document}